# TCAD Cryogenic Model Parameters Calibration based on 65 nm NMOS Experimental Data

Wesley Xi, Vikram Magendar, Alhaji A Sharka, and Hiu Yung Wong, *Senior Member, IEEE*

***Abstract*—In this paper, 65 nm n-type Metal-Oxide-Semiconductor Field-Effect-Transistors (nMOSFETs) are taped out and measured at 292 K to 4.5 K. Technology Computer-Aided-Design (TCAD) model parameters are then calibrated to fit the simulation curves to experimental electrical curves. Transmission electron microscopy (TEM) is used to unveil the dimensions of the transistors to ensure accurate modeling. The experimental and simulation capacitance-voltage (CV) curves, drain current - gate voltage ($I_DV_G$) curves, and drain current - drain voltage ($I_DV_D$) curves are fitted well using the calibrated parameters. The major models calibrated are the Lucent model, which is comprised of an extended PhuMob model, Lombardi model for surface mobility, and the Hänsch high field mobility model. A band-tail model with adjusted band width is applied to fit the sub-threshold swings at all temperatures. The calibrated parameters are expected to be applicable to other technologies because TCAD models are not technology-specific. It is also found that, except for saturation velocity, no new temperature-dependent model is needed. The calibration process and fitting strategies are detailed and are expected to be applicable to calibrate other transistor types.**

***Index Terms*— Band-tail states, Calibration, Cryogenic CMOS, Technology Computer-Aided Design (TCAD)**

## I. Introduction

Cryogenic CMOS is promising for large-scale quantum computers [1], more efficient and reliable deep space systems [2], and data center applications with faster memory access [3], thanks to its improved current vs. leakage ratio, sub-threshold swing, and transconductance [4][5]. One of the cost-effective ways to study the electrical behaviors of cryogenic CMOS is to perform Technology Computer-Aided Design (TCAD) simulations. TCAD has become an integral method for broader semiconductor manufacturing since it dramatically speeds up the setup of new processes, reduces human induced errors, and allows for gathering data that are difficult to obtain experimentally [6].

Modeling the electrical parameters of cryogenic CMOS, however, has been historically challenging due to, firstly, numerical implementation difficulties in certain quantities like carrier density [7]. In a simplified 3D n-doped MOS silicon nanostructure, the electron density can vary by over 2 orders of magnitude per nm, when the electric field is at ~5 MV/m at 15 K [8]. Dramatic changes like this leads to solutions with sharp discontinuities in the electric fields and carrier densities, degrading modeling convergence [8]. The second challenge of cryogenic modeling is the novel physics, such as the partial ionization and the individual effects of several scattering mechanisms [7][9], and the existing room-temperature models do not automatically apply at cryogenic temperatures. Recent TCAD modeling papers have calibrated incomplete ionization models for CMOS down to 20 K [9]. Although there are studies that deconvolved the contributions of phonons, surface roughness, and Coulomb scattering in various types of nMOSFETs and pMOSFETs down to 4.2 K [10][11], these results have not been applied to TCAD models. Wide-temperature-range mobility models have also been developed for bulk MOSFETs in [12] and for Bipolar Junction Transistors (BJT) in [13]. However, models required to model MOSFETs are still lacking.

Another recent discussion centers on the non-ideal effects of sub-threshold slope (SS) at cryogenic temperatures. The Boltzmann limit predicts a linear decrease of SS against temperature in FETs, but SS is experimentally observed to saturate at between 7~15 mV/dec across multiple transistor technologies and structures at both low and high drain voltages when $T < 50$ K [14][15]. The origin of this saturation is debated in literature, with some attributing it to the presence of a band-tail [14][15][16], and while others believe it is caused by traps at the oxide/channel interface [17][18]. Successful TCAD convergence has been demonstrated using the second theory on a n-type MOSFET at 4K [18]. However, it uses an older 0.35 μm technology and does not include quantization models which could worsen convergence but is important on devices with a thinner gate insulator, and it did not simulate a device that matches exactly with the fabricated sizes of the measurements since that information was unknown. While it is possible to simulate cryogenic behavior using *ab initio* calculations [16][19], it is slow and TCAD simulation is preferred.

In this paper, we fabricated test structures on a 65 nm foundry technology, performed CV, $I_DV_G$, and $I_DV_D$ measurements at 300 K, 77 K, and 4.5 K, and calibrated a unified TCAD physics model and parameter sets for nMOSFETs based on the data from 300 K to 4.5 K. Transmission electron microscopy (TEM) is used to reveal the structure of the transistor being measured. Besides presenting the best-fitting models and parameters, the physics, fitting experience, and convergence behavior at cryogenic temperatures are also discussed. To help readability, in the fitting sections, the best results are presented first followed by fitting methods and corresponding physics.

This material is based upon work supported by the NSF under Grant No. 2046220. Wesley Xi, Vikram Magendar, Alhaji A Sharka, and Hiu Yung Wong are affiliated with San Jose State University, San Jose, CA 95192 USA. (Corresponding author: Wong; e-mail: hiuyung.wong@sjsu.edu).

## II. Experiment

Two test structures are taped out and fabricated in a commercial foundry 65 nm RF process [20]. One is a large-area nMOSFET (standard device) with width (W) = 100 µm and length (L) = 20 µm, implemented using a two-finger structure (each has W = 50 µm). Therefore, it has a total area of 2000 µm$^2$. This structure is used for accurate CV measurements due to its large area. Another is a minimal-size nMOSFET (standard device) with W/L = 120 nm/60 nm. This is used for $I_DV_G$ and $I_DV_D$ measurements.

The CV (gate capacitance – gate voltage) measurement is only performed at 300K on the large transistor at $f_1 = 1$ MHz and $f_2 = 100$ kHz with an amplitude of 25 mV using a Keysight E4980A LCR. The gate voltage is swept from -1 V to 1 V with $V_D = V_B = V_S = 0$. For such a large area transistor in 65 nm technology node, the gate leakage is not negligible during CV measurements and correction is required to obtain the correct capacitance, $C_{corr}$. This is done by measuring at the two frequencies, $f_i$, for $i = 1, 2$, and $C_i$ and $D_i$ are corresponding measured capacitance and dissipation factor.

The model used in the measurement is $C_i$ in parallel with an $R_i$, i.e., $C_i \parallel R_i$. The measured dissipation factor $D_i$ is defined as the ratio between the real and imaginary components of the system admittance $Y_i$:

$$Z_i^{-1} = Y_i = G_i + j\omega_i C_i, \text{ (1a)}$$

$$D_i = \frac{G_i}{\omega_i C_i}, \text{ (1b)}$$

where the conductance $G_i$ is the reciprocal of $R_i$. Substituting $G_i$ as the product of $D_i$ and $\omega_i C_i$, the impedance $Z_i$ can be rewritten as:

$$Re\{Z_i\} = \frac{D_i}{\omega_i C_i (1+{D_i}^2)}, \text{ (1c)}$$

$$Im\{Z_i\} = -\frac{1}{\omega_i C_i (1+{D_i}^2)}, \text{ (1d)}$$

In reality, however, a series access resistance $R_s$ also needs to be considered in addition to the parallel gate leakage $R_p$ with a true capacitance, $C_{corr}$ and this represents the true circuit. The corrected impedance $Z_{corr}$ expressing in this true circuit is:

$$Z_{corr} = R_s + \frac{1}{\frac{1}{R_p} + j\omega_i C_{corr}}, \text{ (1e)}$$

Separating the real and imaginary components of $Z_{corr}$, we have,

$$Re\{Z_{corr}\} = R_s + \frac{R_p}{1+{\omega_i}^2 {C_{corr}}^2 {R_p}^2}, \text{ (1f)}$$

$$Im\{Z_{corr}\} = \frac{-\omega_i C_{corr} {R_p}^2}{1+{\omega_i}^2 {C_{corr}}^2 {R_p}^2}, \text{ (1g)}$$

Eq. (1g) is the imaginary part of the true circuit. Eq. (1d) is the imaginary part measured by the equipment and expressed in the simplified $C_i \parallel R_i$ circuit. They should be the same. Therefore, equating Eq. (1d) and Eq. (1g), we obtain:

$${\omega_i}^2 C_{corr} = {\omega_i}^2 C_i (1 + {D_i}^2) - \frac{1}{{R_p}^2 C_{corr}}, \text{ (1h)}$$

Finally, substituting $i = 1, 2$ to form two equations and finding the difference, we get,

$$C_{corr} = \frac{f_1^2 C_1 (1+D_1^2) - f_2^2 C_2 (1+D_2^2)}{f_1^2 - f_2^2}, \text{ (1)}$$

which is the equation used for correction. Fig. 1a shows the measured and corrected CV curves. Although there is a glitch after correction due to noise, it will not affect the TCAD fitting, as the fitting is based on the overall shape.

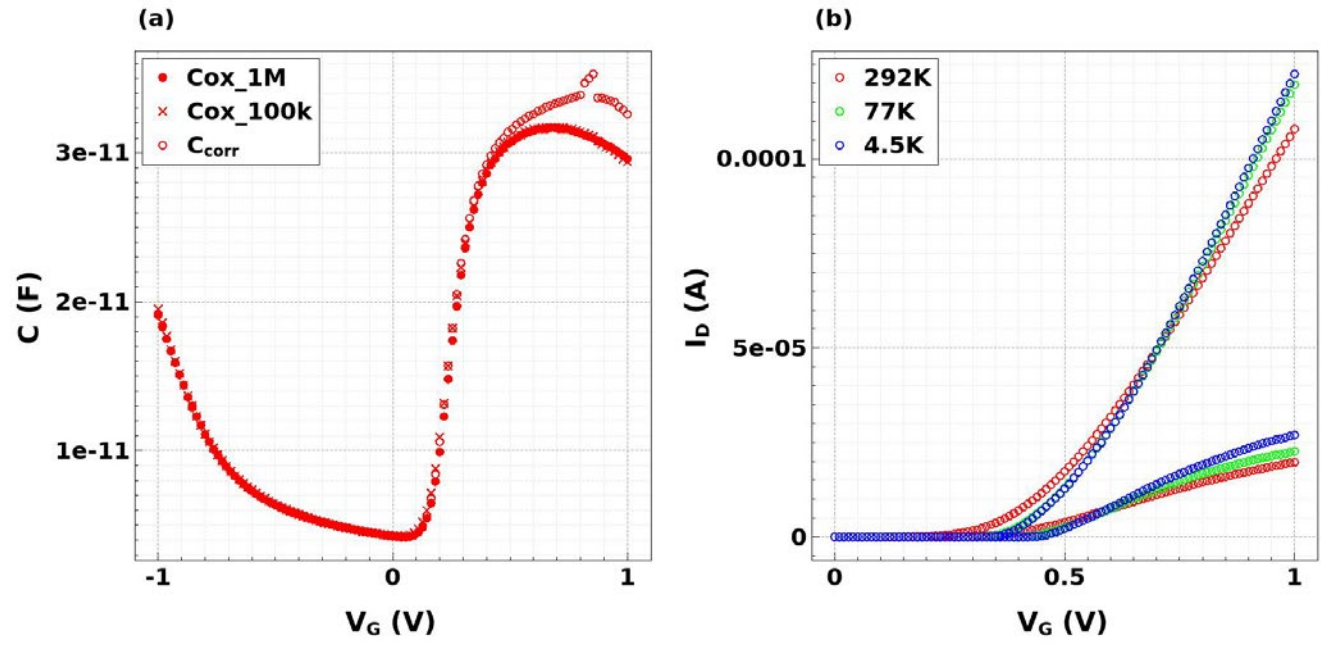

Fig. 1. The measured CV curve (A) at 300 K and $I_DV_G$ curves (B) at various temperatures at $V_D = 1$ V and $V_D = 0.05$ V.

The $I_DV_G$ and $I_DV_D$ curves are measured at 292 K, 77 K, and 4.5 K using a Lakeshore TTPX probe station system with a Keysight E5262A IV analyzer. The samples were cooled down by liquid helium with a temperature controller and heater. To enhance thermal contact, vacuum grease is applied between the backside of the dies and the stage. Measurements are performed 30 minutes after the stage has reached the target temperature. $I_DV_G$ curves are measured from $V_G = 0$ to 1 V at $V_D = 50$ mV and 1V. $I_DV_D$ curves are measured from $V_D = 0$ to 1 V at $V_G =$ 0 V, 0.2 V, 0.4 V, 0.6 V, 0.8 V, and 1 V. Fig. 1b shows the measured $I_DV_G$ curves at $V_D = 50$ mV and 1 V at various temperatures.

## III. Structure and TEM Images

In order to fit TCAD parameters appropriately, it is important to obtain the structural information of the device being measured. Therefore, transmission electron microscopy (TEM) is used to reveal certain critical structural parameters. This is because, firstly, the foundry does not provide details of the structural parameters and, secondly, there are significant device-to-device variations for minimum-size transistors [20]. Fig. 2 shows the TEM. The gate length is measured to be 39 nm. The left and right oxide liners are measured t

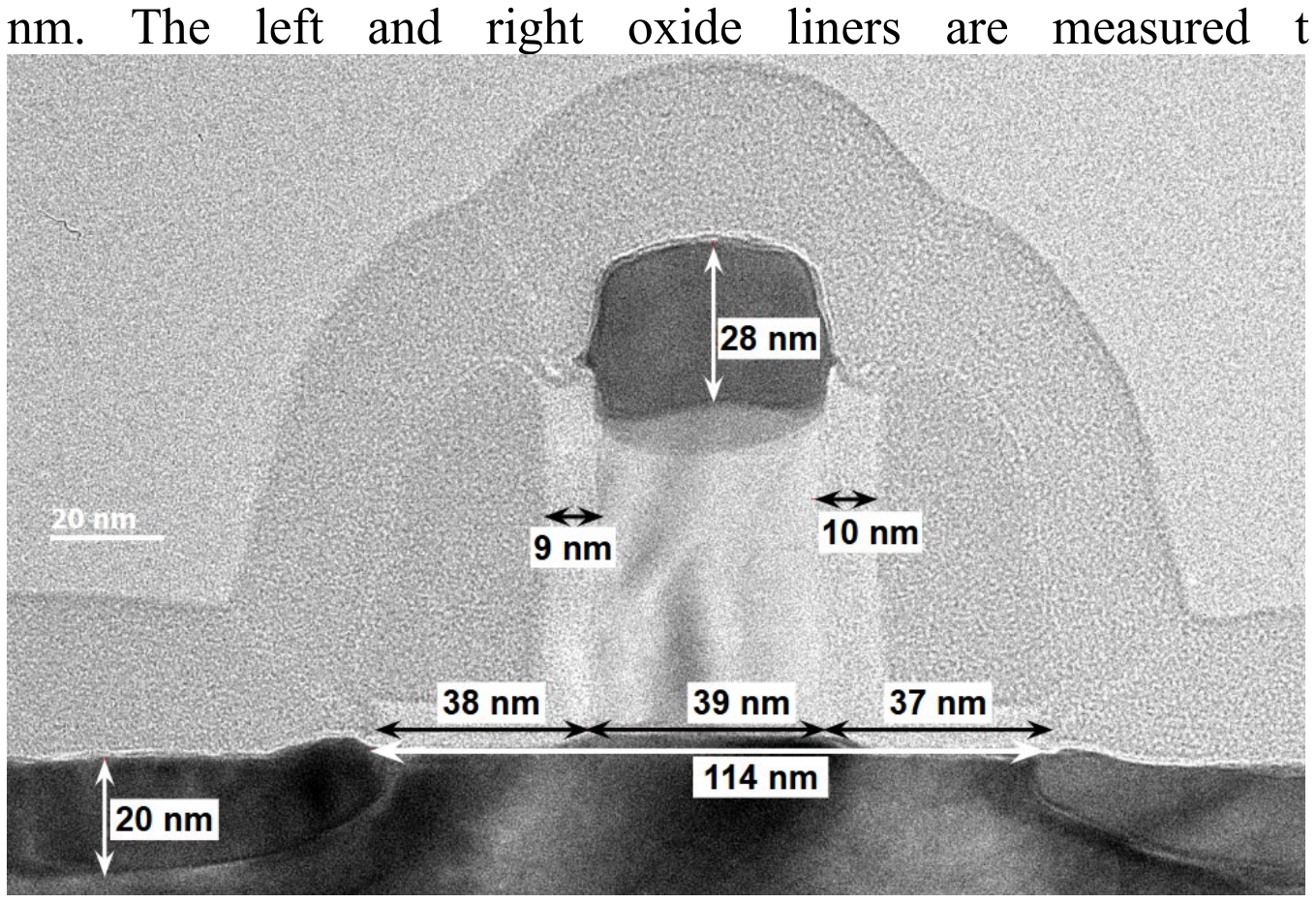

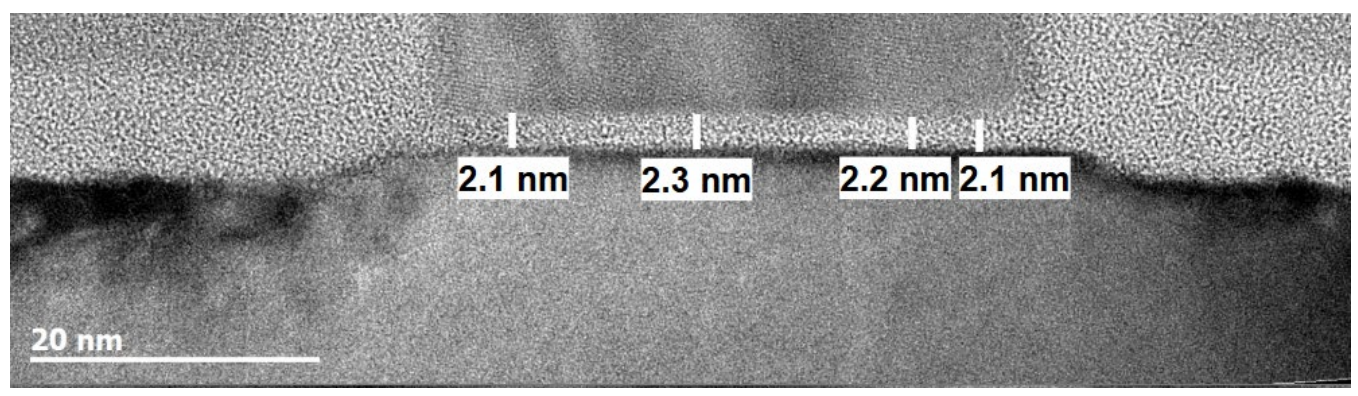

Fig. 2. Top: TEM image of the 65nm transistor. Bottom: TEM in the gate region.

o be 9 nm and

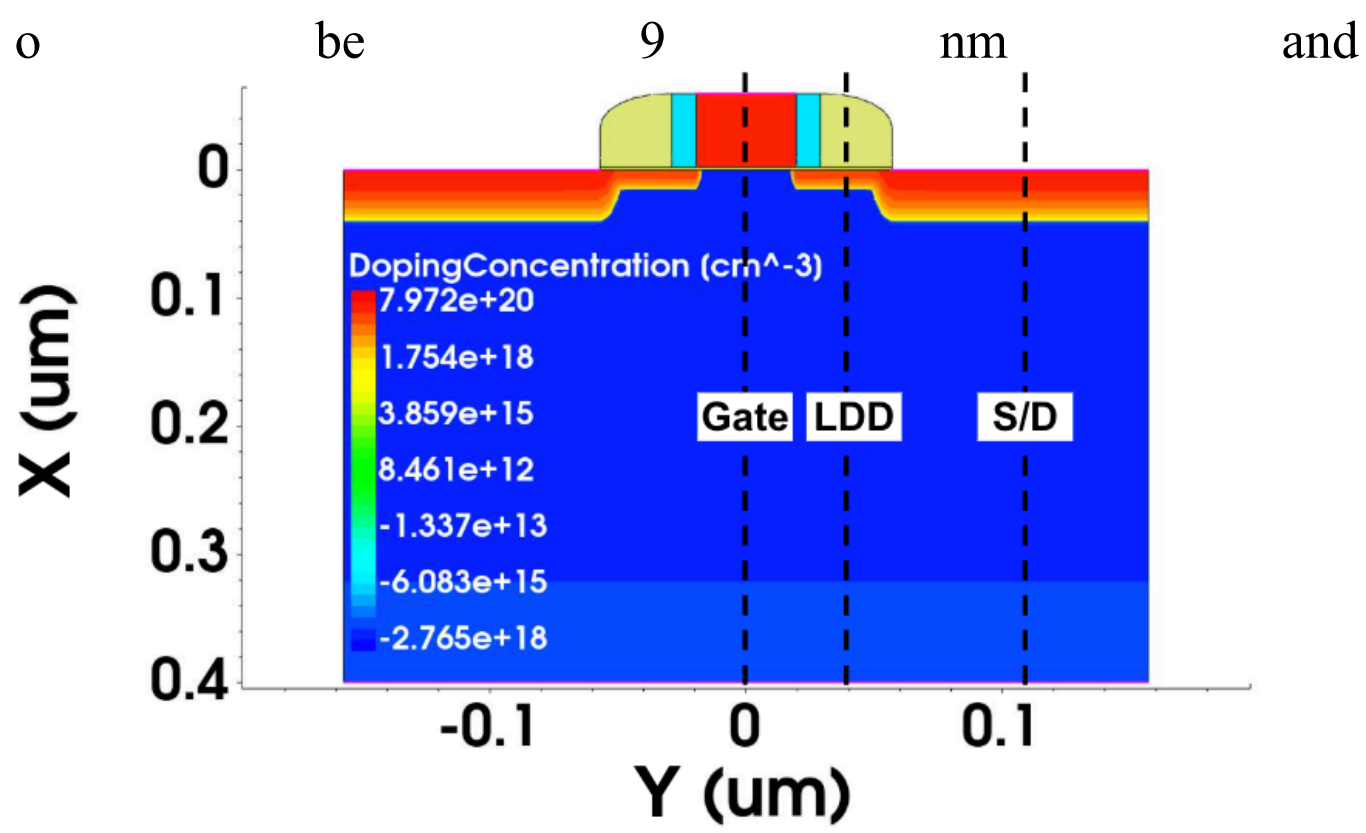

Fig. 3. Cross-section of the simulated 65 nm technology transistor. Note that this is the structure for IV simulation. The structure for CV simulation has L = 20 um without halo doping.

10 nm thick, respectively. The total spacer widths (including the liners) are measured to be 38 nm and 37 nm, respectively. The spacer material is believed to be nitride, which is commonly used in 65 nm technology nodes. The depth of the source/drain silicide is measured to be at least 20 nm. The poly gate silicide is measured to be 28 nm thick. Fig. 2 also shows the TEM image zoomed in on the gate oxide. Four measurements of the gate oxide layer thickness are taken: 2.1 nm, 2.3 nm, 2.2 nm, and 2.1 nm. These measurements were then used to inform the structure creation in the TCAD simulation. Fig. 3 shows the structure created in TCAD Sentaurus Structure Editor.

## IV. Physics Model Selections

We first summarize and discuss the TCAD models necessary to model 65 nm devices from 300 K to 4.5 K. This follows from the fitting process, which will be presented in the following sections. Readers may go directly to the following sections if they are only interested in the fitting results.

*General Models:* We first use the default parameter set provided in TCAD Sentaurus as a starting point, using the keyword DefaultParametersFromFile [21]. A 2D simulation is performed for both transistors using the correct 2D cross-section in the source/drain direction, with terminal current and charge scaled to the actual width. Fermi statistics is used. For example, the electron ($n_B$) density is given by:

$$n_B(E_{F,n},T) = N_C F_{1/2}\left(\frac{E_{F,n}-E_C}{k\mathrm{T}}\right), (2)$$

where $k$ is the Boltzmann constant, $T$ is the temperature, $N_C$ is the effective density of states (DOS) of the conduction band, $E_{F,n}$ is the quasi-Fermi energy, and $E_C$ is the conduction band edge. $F_{1/2}$ is the Fermi integral of order $1/2$. For bandgap narrowing $E_{bgn}$, the default BennettWilson model is used [22]. Both effective bandgap $E_{g,eff}$ and electron affinity $\chi$ are affected by $E_{bgn}$ and they are temperature dependent. The default temperature-dependent parameters are used.

*Band-tail States:* It is widely known that the sub-threshold swing (SS) of all kinds of transistors suffers from the saturation effect, where SS does not scale with $T$ when $T$ is less than about 70 K [10][14][16][17][23], which is believed to be due to additional conduction through the band-tail states [14][16]. It might also be modeled by including band-edge traps at the insulator/silicon interface [17][18]. In this paper, we model the abnormal SS by using band-tail states. This model is only included in the $I_DV_G$ and $I_DV_D$ simulations because CV measurement is only performed at 300 K. The DOS at energy, $\varepsilon$, due to the band-tail states are modeled as,

$$D(\varepsilon) = \frac{N_t}{\sigma\sqrt{2\pi}}\exp\left[-0.5\left(\frac{\varepsilon-\varepsilon_0}{\sigma}\right)^2\right], \quad (3)$$

where $N_t$ is the integral of the band-tail DOS over the energy, $\varepsilon_0$ is the position of the maximum energy in reference to the band edges, and σ is the energy width of the band-tail distribution [21]. As a result, the additional electron density due to the band-tail states, $n_{bt}$, is,

$$n_{bt}(E_{F,n},T) = \int_{-\infty}^{\infty}\frac{D(\varepsilon)}{1+\exp\left(\frac{\varepsilon+E_C-E_{F,n}}{kT}\right)}d\varepsilon. \quad (4)$$

The total election density, $n$, is therefore,

$$n(E_{F,n},T) = n_B(E_{F,n},T) + n_{bt}(E_{F,n},T). \quad (5)$$

*Quantum Correction:* To model the quantization effect, which increases threshold voltage and reduces the gate capacitance [21], a potential-like quantity $\Lambda_n$ is used in Synopsys SDevice. It modifies Eq. (2) to be:

$$n = N_C F_{1/2}\left(\frac{E_{F,n}-E_C-\Lambda_n}{kT_n}\right). \quad (6)$$

We have selected the density gradient model [25][26] to calculate $\Lambda_n$ in the silicon region, and the density-based formula instead of the potential-based formulas is selected for better convergence with

$$\Lambda_n = -\frac{\gamma\hbar^2}{6m_n}\frac{\nabla^2\sqrt{n}}{\sqrt{n}}, (7)$$

where γ is a fit factor, $\hbar$ is the reduced Planck constant, and $m_n$ is the effective mass of the electron. It is worth noting that the homogeneous Neumann boundary condition is used (i.e., $\hat{n}\cdot\nabla\Lambda_n^+ = 0$, where $\Lambda_n^+$ is in the region of the interface where the density gradient equation is not solved, e.g., the insulator). This is valid when the quantum effects are insignificant [21], as in our case, due to the relatively thick gate insulator and it is found to match the experimental result better.

*Mobility:* The Lucent model [27] and its parameters are used to model the mobility in silicon. The Lucent model is comprised of three parts: an extended Klaassen's PhuMob model [28], Lombardi model for surface mobility [29] with adjusted parameters, and the Hänsch high field mobility model [30]. The electric field parallel, $F_{hfs}$, to the insulator/silicon interface is used as the driving force for high-field mobility calculation. To avoid double-counting, the constant phonon scattering mobility model is disabled, as PhuMob includes phonon scattering. The low field mobility, $\mu_{low}$, is given by,

$$\frac{1}{\mu_{low}} = \left(\frac{1}{\mu_L}+\frac{1}{\mu_{DAeh}}\right) + \left(\frac{1}{\mu_{ac}}+\frac{1}{\mu_{sr}}\right), \quad (8)$$

where $\mu_L$ and $\mu_{DAeh}$ are the PhuMob terms and are the lattice scattering mobility and other bulk scattering mechanisms, respectively. The details of their expressions may be found in [21] or [27]. $\mu_{ac}$ and $\mu_{sr}$ are the acoustic phonon scattering and the surface roughness scattering of the Lombardi model, respectively. $\mu_{ac}$ is given by,

$$\mu_{ac} = \frac{B}{F_\perp} + \frac{C((N+N_2)/N_0)^\lambda}{F_\perp^{\frac{1}{3}}\left(\frac{T}{300K}\right)^k}, (9)$$

where $N$ is the total doping concentration, $F_\perp$ is the normal component of the electric field at the interface, $T$ is the temperature, and B, C, $\lambda$, and $k$ are fitting parameters. $N_2 =$ 1cm$^{-3}$, and $N_0 =$ 1cm$^{-3}$ are used to match the equation in [27]. $\mu_{sr}$ is given by

$$\mu_{sr} = (\frac{(F_{\perp}/F_{ref})^{A^*}}{\delta} + \frac{F_{\perp}^3}{\eta})^{-1}, (10)$$

where $F_{\perp}$ is the vertical electric field, $F_{ref}$ is 1 V/cm for normalization, and $\delta$ , $\eta$ , and $A^*$ are fitting parameters, respectively. The high field saturation mobility is given by,

$$\mu(F) = \frac{2\mu_{low}}{1+[1+4\left(\frac{\mu_{low}F_{hfs}}{v_{sat}}\right)^2]^{\frac{1}{2}}}, (11)$$

where $v_{sat}$ is the saturation velocity [27]. This combination is similar to a previous paper published by our group [18], which uses the original PhuMob and Lombardi models and parameters. The Lucent model is used in this paper because it offers better physical explanations. The original Lombardi model uses the empirical Masetti bulk mobility model [32], which, according to the Lucent paper [27], has limited physics since it accounts for bulk Coulomb scattering only as a function of doping density but not carrier density. The reason Lombardi does this is to save on computation costs, but it is not physical since more free carriers can weaken impurity scattering. Instead, Lombardi's model ends up absorbing the dependence of carrier density to the $\lambda$ parameter in Eq. (9). Lombardi notes himself that this is unphysical as phonon scattering should not depend on carrier density. In comparison, Lucent uses the PhuMob bulk model, which has better Coulomb scattering physics that accounts for both doping and carrier density screening. This means that the Lucent acoustic phonon scattering parameters do not have to compensate for bulk Coulomb scattering as much as the original Lombardi model does. In addition, Lucent accounts for the dependence of surface roughness scattering on both inversion charge density and electric field, while the original Lombardi model only accounts for the electric field.

Compared to the original PhuMob, the PhuMob in the Lucent model does not have ultra-high carrier concentration scattering or the electron-hole scattering terms. Disabling electron-hole scattering is reasonable since the electron-hole scattering should not have a big effect on nMOSFET, where the carriers are dominated by electrons.

The Hänsch model in Eq. (11) is an extended version of the Canali model [33]. The reason we select the Hänsch model instead of the default Canali model is that it has better convergence.

It should also be noted that the bulk low-field and high-field mobility developed in [12] is not used because implementing it in commercial software is difficult for users and convergence may be hindered by negative differential mobility.

*Stress:* The stress model is also turned on in the $I_DV_G$ and $I_DV_D$ simulations. Since the transistor for CV measurement has a large W and L, the stress effect is likely negligible. Therefore, stress is not turned on in the CV simulations. The deformation potential model is activated. The $k \cdot p$ method for electron bands is used to account for shear strain components. Minimum band energies are used to compute the conduction and valence band edges. The stress-dependent mobility model for electrons is activated. Several options are turned on within this mobility model: Fermi statistics carrier concentration dependency, interface scattering, and stress-induced change of the electron effective mass, with the electron saturation factor set to 0. Stress is set to 1 GPa.

*Other Models:* In the recombination section of silicon physics, we have used the constant carrier generation model at a rate of $10^{10}$cm$^{-3}$ to improve convergence at cryogenic temperature, like how it is done in ultra-wide bandgap device simulations [31]. The Shockley-Read-Hall generation is also turned on. Incomplete ionization is not turned on in the simulation because dopants are fully ionized in the inversion and depletion region during IV and CV simulations. The lightly-doped drain region (LDD) and the source/drain (S/D) region are expected to form impurity bands due to their high concentrations, obviating the need to turn on the incomplete ionization model [9].

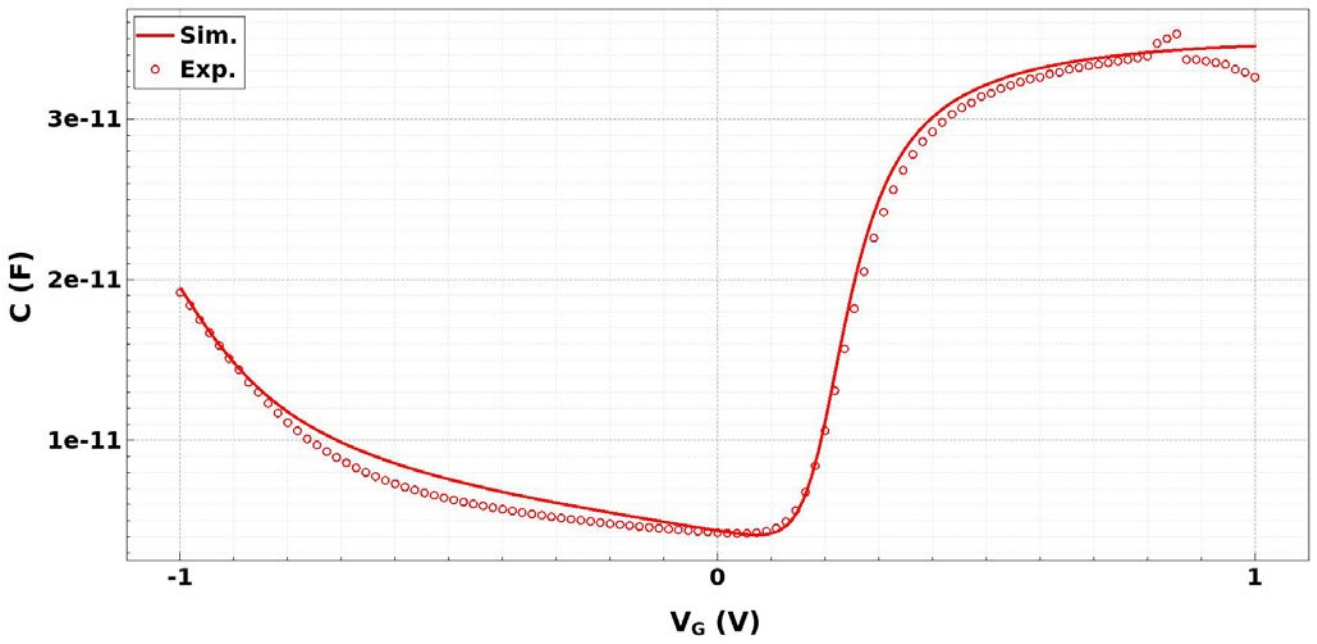

Fig. 4. The comparison of TCAD CV simulation and the experimental CV.

## V. CV Fitting

The purpose of CV fitting is to obtain the correct substrate doping profile and gate stack parameters. Since it has a large W and L, the stress effect, LDD, and S/D profiles have a negligible effect on its fitting. Moreover, halo implants, which modify the substrate doping only near the S/D, are also expected to have minimal effect on CV measurements. In the following discussion and simulation, only the gate stack and substrate doping are discussed; the simulation uses the same setup as the IV simulation in the following section, except without halo doping.

Fig. 4 shows the comparison between the CV simulation and the experiment. The simulated result matches closely with the measurement. The substrate doping is simplified to be a combination of a constant p-type background profile at 1×10$^{15}$ cm$^{-3}$ and a Gaussian profile in the substrate region with $n_{peak1}$ and $x_{depth1}$ being the peak value and depth, respectively. The location of the peak ($n_{peak1}$) is defined as $x_{peak1} = 0$ nm from the insulator/silicon interface. $x_{depth1}$ is defined as the distance at which the concentration becomes $0.1 \times n_{peak1}$ . Gate insulator thickness, $t_{ins}$, and its dielectric constant, $\varepsilon_{ins}$, are also adjusted to fit the experimental CV. Table I shows the fitting strategy. Each parameter is adjusted to fit the corresponding target. It can be seen that the maximum capacitance ($C_{max}$) increases as $t_{ins}$ drops. Note that gate depletion and quantum correction are modeled in CV fitting.

TABLE I
CV Fitting Strategy

| Steps | Parameter | Target | Relationship |
|---|---|---|---|
| 1.1 | $t_{ins}$ | $C_{max}$ | $t_{ins}$ ↗, $C_{max}$ ↘ |
| 1.2 | $\varepsilon_{ins}$ | $C_{max}$ | $\varepsilon_{ins}$ ↘, $C_{max}$ ↘ |
| 2 | $n_{peak1}$ | $V_{th}$ | $n_{peak1}$ ↗, $V_{th}$ ↗ |
| 3 | $x_{depth1}$ | $C_{min}$ | $x_{depth1}$ ↗, $C_{min}$ ↗ |

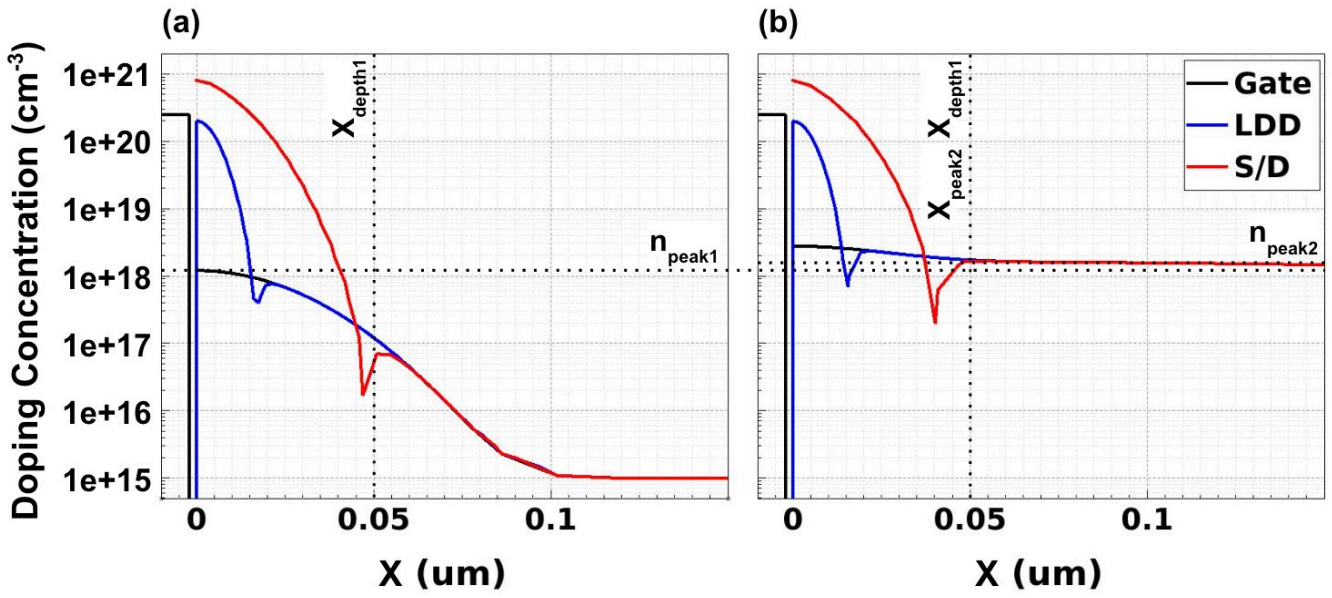

Fig. 5. Doping concentration at various cut locations in Fig. 3 for the CV structure (left) and IV structure (right).

The best fit of maximum capacitance occurred at an equivalent oxide thickness (EOT) of 1.57 nm. Since oxynitride is known to be used in 65 nm, based on the TEM image with an insulator thickness range of 2.1~2.3nm in Fig. 2, a 2.16 nm oxynitride is used as the insulator with a dielectric constant of 5.36. Increasing $n_{peak1}$ shifts the curve to the right (increases the threshold voltage, $V_{th}$) and raises minimum capacitance ($C_{min}$). Decreasing $x_{depth1}$ drops $C_{min}$. The best fit occurs when $n_{peak1}$= 1.2×10$^{18}$ cm$^{-3}$ and $x_{depth1}$ = 50 nm. The parameter values are obtained and shown in Table II. Fig. 5 shows the doping concentration at various cut locations of the CV structure.

TABLE II
CV FITTING PARAMETER VALUES

| Parameter | Fitted Value | Units |
|---|---|---|
| $t_{ins}$ | 2.16 | nm |
| $\varepsilon_{ins}$ | 5.36 | - |
| $n_{peak1}$ | $1.2 \times 10^{18}$ | cm$^{-3}$ |
| $x_{depth1}$ | 50 | nm |

## VI. IV FITTING

Figs. 6, 7, and 8 show the comparison between the $I_DV_G$ and $I_DV_D$ simulations and the experimental measurements. Both logarithmic scale and linear scale are shown for the $I_DV_G$ curves. The simulation curves match closely with the experiments in general. Table III shows the fitting strategy. The fitting is split into three parts. Firstly, $V_{TH}$ and Drain-Induced Barrier Lowering (DIBL) are fitted together. Then, the sub-threshold swing (SS) is fitted. Finally, the on-state current, $I_{on}$, is fitted. It should be noted that the fitting is iterative, and the CV and IV fitting need to be verified at the same time.

TABLE III
IV FITTING STRATEGY

| Steps | Equation | Item | Fitting Target | Relationship |
|---|---|---|---|---|
| 1.1 | - | $n_{peak2}$ | $V_{TH}$ | $n_{peak2}$ ↗, $V_{TH}$ ↗ |
| 1.2 | - | $X_{j,LDD}$ | DIBL | $X_{j,LDD}$ ↘, DIBL ↘ |
| 2 | (3) | $\sigma$ | SS | $\sigma$ ↗, SS ↗ |
| 3 | (10) | A* | curvature | A* ↗, curvier |
| 4.1 | (9) | $C$ | $I_{ON,\,292K}$ | C ↗, $I_{ON,\,292K}$ ↗ |
| 4.2 | (11) | $v_{sat}$ | $I_{ON,\,SAT}$ | $v_{sat}$ ↗, $I_{ON,\,SAT}$ ↗ |
| 5 | (9) | $k$ | $I_{ON,\,77\&4K}$ | k ↗, $I_{ON,\,77\&4K}$ ↗ |

To fit $V_{TH}$, the same substrate doping profile in CV fitting is used. However, a second Gaussian doping profile (with $n_{peak2}$, $x_{peak2}$, $x_{depth2}$) in the substrate is added to account for an effective halo doping on this short and narrow transistor. One also needs to adjust the LDD profile to match the DIBL. However, only 292 K needs to be fitted and it should fit other temperatures relatively well. It is found that, when $x_{peak2}$ = 50 nm, $n_{peak2}$ = 1.6×10$^{18}$ cm$^{-3}$, and $x_{depth2}$ = 550 nm with an effective LDD junction depth ($X_{j,LDD}$) of 15 nm, both saturation and linear $V_{TH}$ are fitted well. Fig. 5b shows the doping concentration cuts of the transistor for IV simulations. It should be noted that the SS of the room temperature measurements are also matched in this process. This is an iterative result of choosing the right gate dielectric, gate doping, substrate doping, and $X_{j,LDD}$.

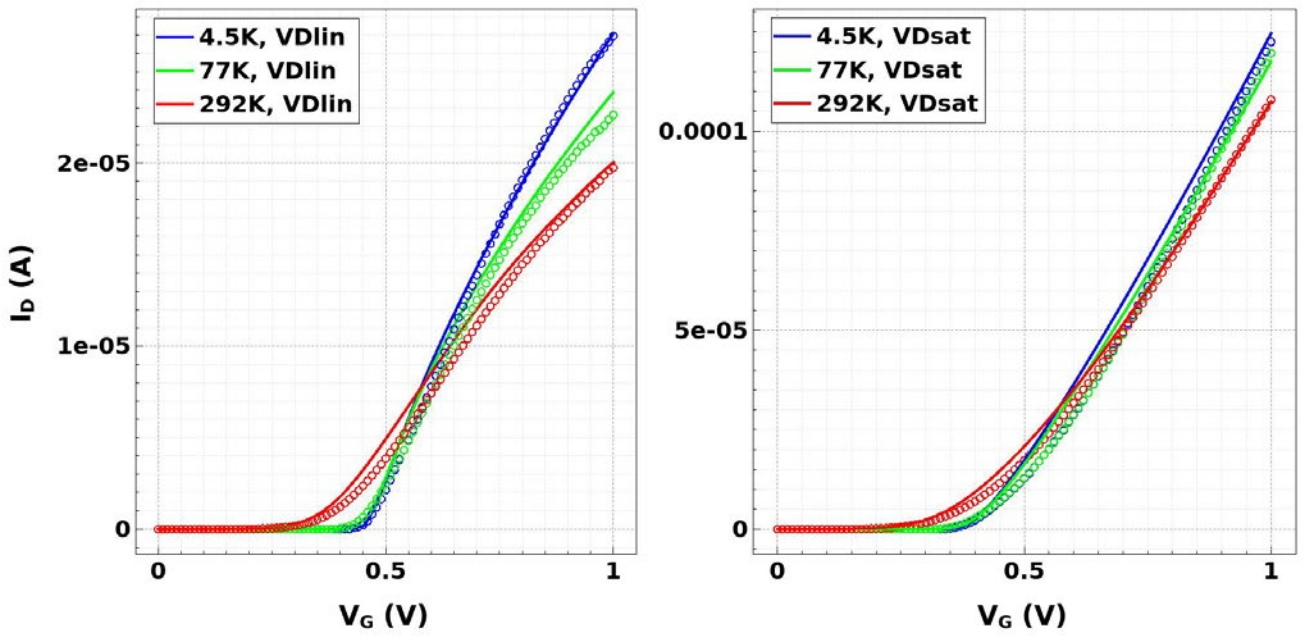

Fig. 6. The comparison of $I_DV_G$ simulations (lines) and the experimental measurements (circles) at $V_D$ = 0.05 V (left) and 1 V (right). The current is in linear scale.

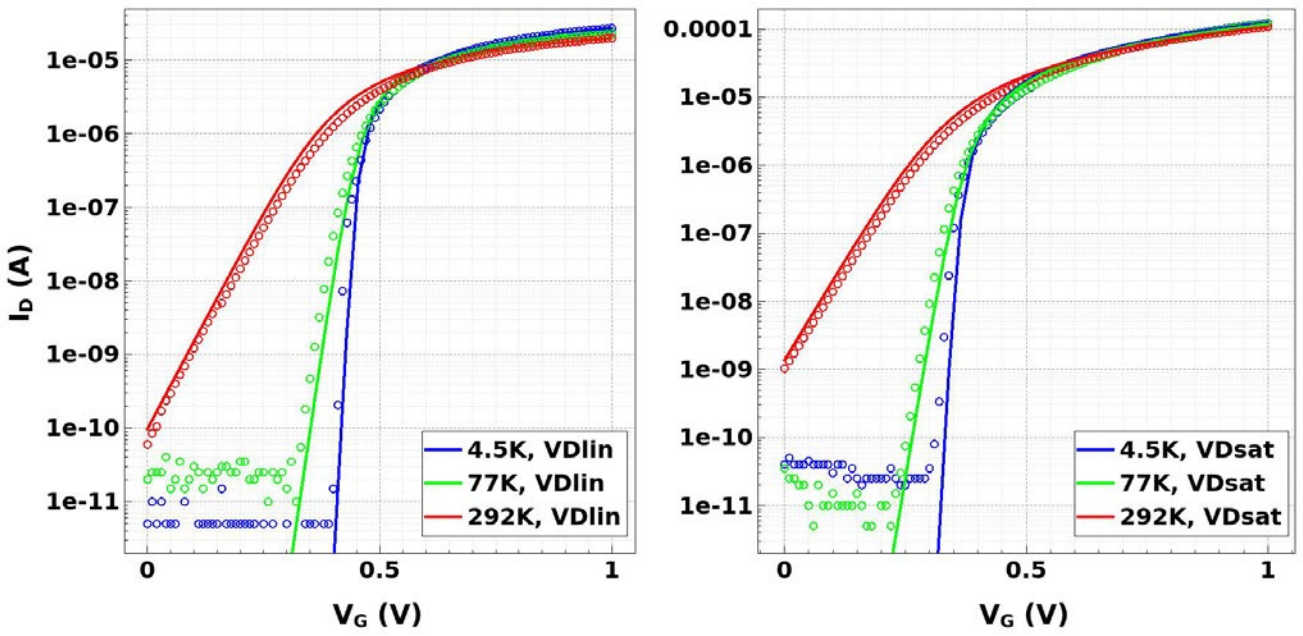

Fig. 7. The comparison of $I_DV_G$ simulations (lines) and the experimental measurements (circles) at $V_D$ = 0.05 V (left) and 1 V (right). The current is on a log scale.

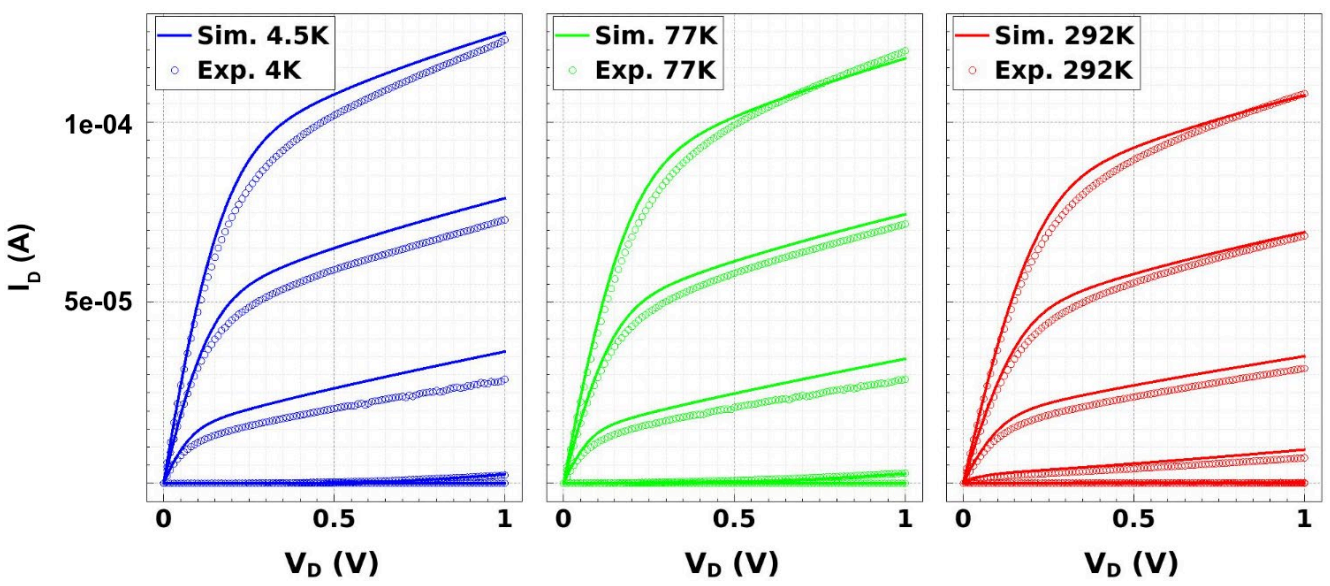

Fig. 8. The comparison of $I_DV_D$ simulation and the experiments at various temperatures. From top to bottom, $V_G$ = 1.0, 0.8, 0.6, 0.4, 0.2, 0.0 V.

To match the SS at cryogenic temperature, the band-tail width $\sigma$ in Eq. (3) is changed. It has no effect on the room temperature fitting. As $\sigma$ increases, 4.5 K SS increases (more gentle) and can match the experimental abnormal SS better. $\sigma = 0.004\ eV$ is found to match the SS of all temperatures well.

The current fitting is achieved by adjusting mobility parameters. To get the low field current fitting, we have focused on adjusting the Lombardi parameters instead of the ones in PhuMob because the Lombardi model accounts for surface roughness, which is process-dependent. The parameter $A^*$ in Eq. (10) affects the curvature of the inversion region:

$$A^* = A + \frac{(\alpha_{\perp,n}n+\alpha_{\perp,p}p)}{(N/N_{ref})^{v}}, (12)$$

where $A$ is the base term in the original Lombardi model [29], and the fraction term is proposed by the Lucent model [27] as the dependence of surface roughness scattering on carrier density. $\alpha_{\perp,\text{n}}$, $\alpha_{\perp,\text{p}}$, and $v$ are fitting. $n$ and $p$ are electron and hole concentrations. $N_{ref} = 1\ cm^{-3}$ is used to cancel the unit of the doping concentration. It is found that the base term $A$ mainly affects the curvature of $I_DV_G$ in the on-state. Decreasing $A$ makes the inversion region less curvy (i.e., less dependent on the vertical electric field), which brings the curves to look closer to the experiment. We have settled on $A = 2.0$. Since the second term in Eq. (10) can be ignored due to its small value, the corresponding $A^*$ falls between Matsumoto's prediction of 2.0 [34] (also in our other experiments [10][11]), and Takagi's experimental fitting of 2.6 [35]. Note that this term is expected to be temperature-insensitive and thus is only fitted with the room temperature data [10][11].

Finally, we use the parameter $C$ and temperature dependence $k$ in acoustic phonon scattering in Eq. (9), as well as saturation velocity $v_{sat}$ in Eq. (11), to get a close current fit at both $V_D$ = 50 mV and 1 V at all three temperatures. The strategy is first using $C$ to get a good fit at $V_D$ = 0.05 V, 292 K, and then using $v_{sat}$ to get a good fit at $V_D$ = 1.0 V, 292 K. Although $C$ primarily affects the low field mobility and $v_{sat}$ primarily affects high field, they need to be adjusted together, since fitting one $V_D$ condition has the side effect of causing the other to shift with it. The final step is taken to fit the current at both 77 K and 4.5 K and at both $V_D$ with the parameter $k$. This is similar to what another paper did that was published by our group [18]. Decreasing either $C$ or $k$ worsens the inversion region mobility and lowers the on-state current. The final values we have adopted are $C$ = $5.3\times10^3$ $cm^{5/3}V^{-2/3}s^{-1}$, $k$ =0.01, and $v_{sat}$ is described as:

$$v_{sat} = v_0 \times \sqrt{\tanh\left(\frac{T_c}{T}\right)}, (13)$$

where $v_0 = 1.74\times10^7$m/s and $T_c$ = 332K. It is worth noting that our fitted $k$ value is very small compared to [27], which indicates a weak temperature dependence of the acoustic phonon scattering between 4.5K and 292K. This result is similar to [18], where $k$ is also fitted to be lower than expected. Table IV shows the final device and model parameters used to fit IV curves at various temperatures in addition to those in Table II, compared to literature values in [27] (the values with asterisks are default values in the simulator).

## VII. Discussions

### A. Sub-threshold Slope Behaviors

In the fitting process, it was found that SS is affected by two models. The first is the band-tail model. Fig. 9 shows the effects of band width $\sigma$ in Eq. (9) on SS. SS is defined as the voltage difference between current at $5\times10^{-9}$ A and $5\times10^{-8}$ A. The non-ideal effect of SS flattening out is observed when the temperature is low. As $\sigma$ gets larger, this non-ideal effect can also be seen at higher temperatures. The second model that is seen to affect SS is the density gradient model in Eq. (7). Fig. 10 shows SS against temperature at different $\sigma$ when the density gradient model is turned off. Compared to Fig. 9, SS drops by roughly 3mV/dec at higher temperatures and at lower $\sigma$. This is likely because the density gradient model affects the EOT and thus the SS. Therefore, in the fitting process of SS, the density gradient model parameters need to be closely monitored.

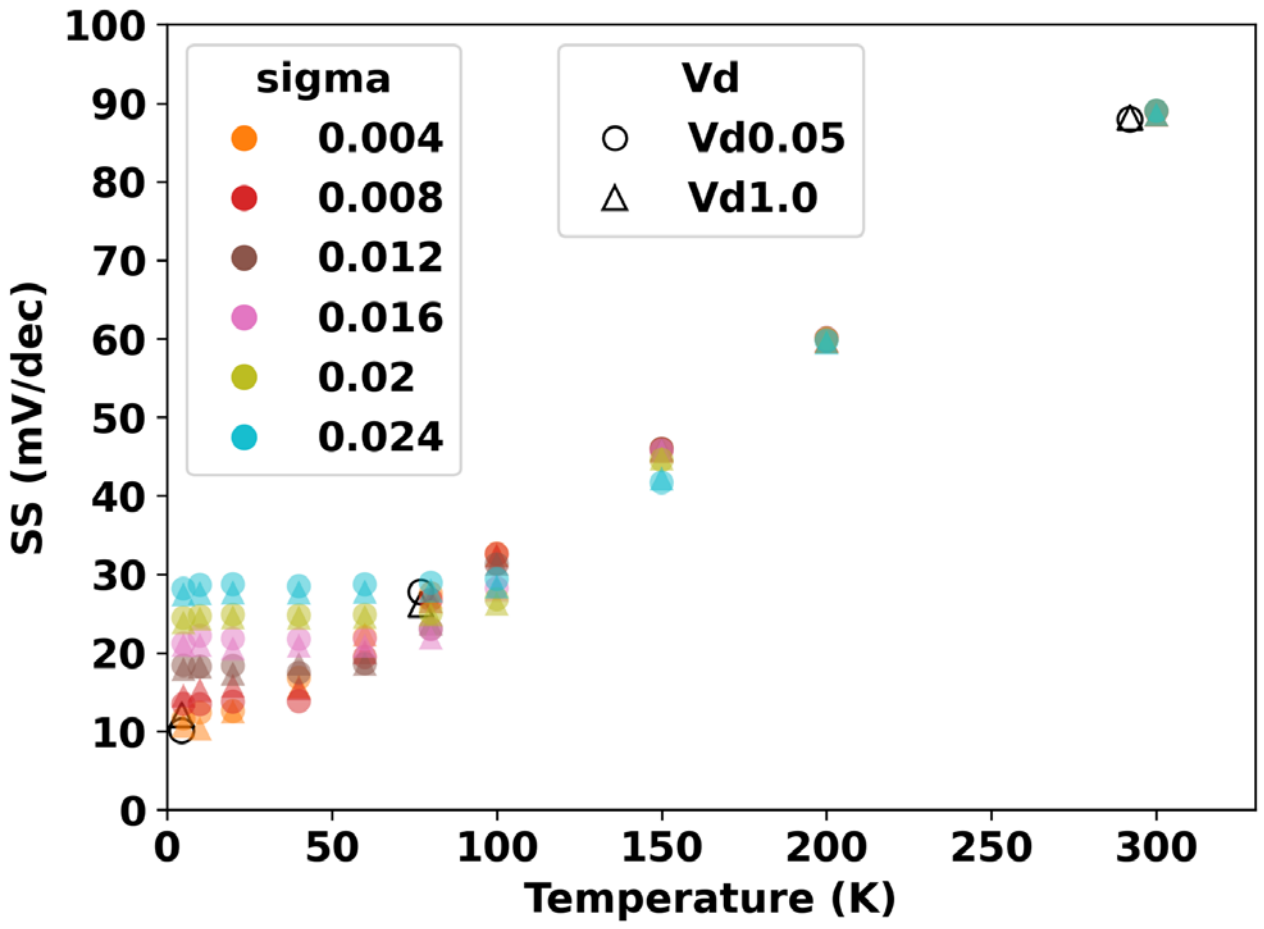

Fig. 9. SS against temperature with different $\sigma$. Quantum potential model is turned on. The hollow circles and triangles are extracted from the 65 nm transistor measurements.

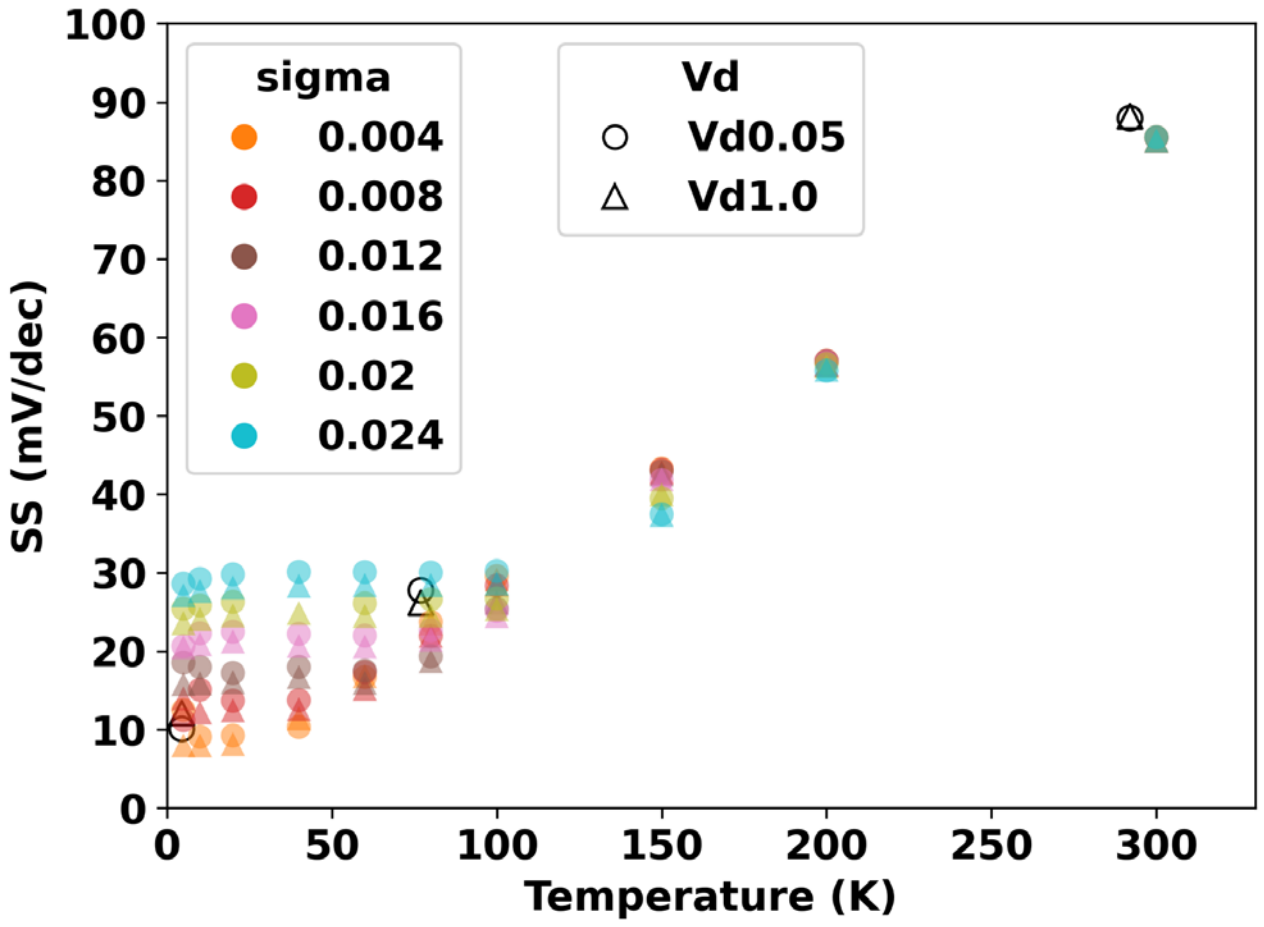

Fig. 10. SS against temperature with different $\sigma$. Quantum potential model is turned off. The hollow circles and triangles are extracted from the 65 nm transistor measurements.

### B. Low-T Phonon Scattering Temperature Dependence

In the literature, some measurements [36][37] suggest that the temperature exponent, $\theta$, in $\mu_L = \mu_{max}(\frac{T}{300K})^{-\theta}$ and $k$ in acoustic phonon scattering, $\mu_{ac}$, in Eq. (9) are temperature dependent, especially at $T$ < 150 K. However, in this study, good fitting was obtained by assuming them to be temperature independent. Therefore, the basic temperature-dependent relationships in all equations are largely maintained in the fitting process, except for the introduction of Eq. (13) and the parameters in Table IV are all temperature-independent.

TABLE IV
IV FITTING PARAMETER VALUES

| Equation | Item | Literature Value | Fitted Value | Units |
|---|---|---|---|---|
| - | $n_{peak2}$ | N/A | $1.6\times10^{18}$ | $cm^{-3}$ |
| - | LDD | N/A | 15 | nm |
| (3) | $\sigma$ | *0.02 | 0.004 | eV |
| (9) | C | $1.7\times10^{4}$ [27] | $5.3\times10^{3}$ | $cm^{5/3}V^{-2/3}s^{-1}$ |
|  | k | 1.7 [27] | 0.01 | - |
| (12) | A | 2.58 [27] | 2.0 | - |
| (13) | $v_0$ | *$1.45\times10^{7}$ | $1.74\times10^{7}$ | m/s |
|  | $T_C$ | *155 | 332 | K |

*SIMULATOR DEFAULT VALUES

### C. Notes on Convergence

Throughout the fitting process, several options and techniques are found to help the convergence in the cryogenic simulations. Firstly, all simulations start at T = 292 K. Then, line search damping is set to $1\times10^{-8}$, for the initial Poisson and density gradient equations [21]. Secondly, a quasi-stationary ramp of the transistor is performed to reach $V_G$ = 1.0 V and the corresponding $V_D$ with the continuity equation. Thirdly, a transient sweep of the transistor from 300K down to the target temperature is performed. Finally, $V_G$ is ramped back to 0 V to get the $I_DV_G$. The key to this process is to solve at the on-state first at cryogenic temperatures before ramping down to the off-state, since it is harder to find a solution at the off-state than at the on-state.

The density gradient model can cause convergence problems, and the DirectQuantumCorrection switch is used. Physically, this option applies the solution of quantum potential directly to the electrostatic potential as a quantum-mechanical band edge, instead of the default option where it would be applied as a multiplication factor to carrier concentration [21]. In some simulations, it was recommended to adjust the reference density parameter, which reduces quantization effects at carrier concentrations lower than its set value, to help with convergence especially when used with Avalanche models. However, during this study, we have observed no such benefit, and this parameter in fact has a strong effect on deep sub-threshold transport behaviors.

Using the Hänsch model instead of the default Canali model for the high field mobility is also important for the fitting to converge. The Extended Precision of 80 digits with iterative linear solver (ILS) is also critical to the success of convergence.

## VIII. CONCLUSIONS

A 65 nm NMOS is measured at various temperatures to obtain its CV, $I_DV_G$ and $I_DV_D$ curves. TEM is used to unveil its structure parameters to construct an appropriate TCAD simulation device. A modified PhuMob model using the Lucent parameters, the band-tail model, and the saturation velocity model are then calibrated to fit the experimental $I_DV_G$ and $I_DV_D$ data from 292 K down to 4.5 K and the CV data at 292 K. Simulations can fit the data using a single set of setup and parameters well in CV and IV in both sub-threshold and on-state regimes. Besides the introduction of temperature-dependent saturation velocity, the current TCAD models are enough to fit the data through appropriate parameter adjustments.

## IX. ACKNOWLEDGMENT

This material is based upon work supported by the National Science Foundation under Grant No. 2046220.